\documentstyle[pra,aps,multicol,eqsecnum,epsfig]{revtex}

\def\be{\begin{eqnarray}}
\def\ee{\end{eqnarray}}

\begin{document}

\title{
Reconstruction of motional states of neutral atoms
via {\em MaxEnt} principle
}
\author{Gabriel Drobn\'y$^{1}$ and Vladim\'{\i}r Bu\v{z}ek$^{1,2}$}
\address{ 
$^{1}$Research Center for Quantum Information, 
Slovak Academy of Sciences,
D\'ubravsk\'a cesta 9, 842 28 Bratislava, Slovakia
\\
$^{2}$Faculty of Informatics, Masaryk University, Botanick\'a 68a,
602 00 Brno, Czech Republic 
}

\date{2 January 2002}                                                          
\maketitle


\begin{abstract}
We present a scheme for a reconstruction of states of quantum systems
from incomplete tomographic-like data. 
The proposed scheme is based on the Jaynes principle of  
{\em Maximum Entropy}. We apply our  algorithm 
for a reconstruction of  motional quantum states
of neutral atoms. As an example we analyze 
the experimental data obtained  
by the group of C. Salomon at the ENS in Paris and we reconstruct
Wigner functions of motional quantum states of Cs atoms trapped in
an optical lattice.
\end{abstract}


\section{Introduction}
\label{sec1}

A reconstruction of  states of quantum systems from experimental data
represents an important tool for a  verification of
 predictions of quantum theory. It also allows us to check the fidelity of 
quantum state preparation 
as well to study the fidelity of processing of information encoded
in states of quantum systems.
Complementary,  discrepancies between estimated (reconstructed) states
based on the measured data and theoretical predictions
can serve as 
an indicator of various noise sources which occur during 
the quantum information 
processing or in the measurement of quantum states.
Without {\em a priori} assumptions about the character of physical processes
and properties of reconstruction schemes
the reconstruction cannot distinguish 
between imperfections related to an incoherent  quantum state
processing and non-ideal measurements \cite{Jaynes1963}.   
Determination of limits for coherent control of quantum degrees 
of freedom or identification of sources of decoherence 
are essential for  systems which are considered for quantum computing 
and information processing \cite{Nielsen2000}.

In atomic optics a highly coherent control of motional degrees of freedom 
has been achieved for trapped ions \cite{Wineland98}
and recently also for neutral atoms \cite{Bouchoule99,Morinaga99}. 
Cold atoms can be cooled into specific quantum states 
within  micro-wells of an optical lattice which is induced by laser beams.
Cold neutral atoms in optical lattices represent a promising system
for quantum information processing. 
To verify a degree (fidelity) of coherent control over motional degrees of
freedom of neutral atoms a reconstruction of their motional quantum states 
from measured data has to be considered. 
We develop a reconstruction procedure based on the Jaynes principle of 
Maximum Entropy ({\em MaxEnt}) \cite{Jaynes1957,Fick1990,Kapur1992}
to achieve this goal. 
This scheme allows us to perform a state 
reconstruction from the experimental data obtained at the ENS in Paris. 

In Section \ref{sec2} we present a brief description of the reconstruction
procedure of a density operator of a quantum system based on the
Jaynes principle of Maximum Entropy. In Section \ref{sec3} we utilize
the MaxEnt principle for development of the reconstruction scheme of motional
states of atoms and we perform numerical tests of our approach in Section
\ref{sec3}. Our reconstruction scheme is applied to the experimental data
in Section \ref{sec4}.

\section{MaxEnt principle and reconstruction of density operators}
\label{sec2}
Let us assume 
a set of observables $\hat G_\nu$ $(\nu = 1,\ldots , n)$
associated with the quantum system under consideration. This system
is prepared in an unknown state $\hat{\rho}$. Let us assume
that from a measurement performed over the system meanvalues 
$\overline{G}_\nu$ of the observables  $\hat G_\nu$ are found. 
The task is to determine
(estimate) the unknown state of the quantum system based 
on the results of the measurement. 
Providing the set of the observables $\hat G_\nu$ is not equal
to the {\em quorum} (i.e., the complete set of system observables
\cite{Newton68}), 
 then the measured meanvalues do not determine the
state uniquely. Specifically, there 
is a large number of density operators 
which fulfill the conditions
\be
{\rm Tr} \,\hat \rho _{\{\hat G\}} &=& 1,
\nonumber\\
{\rm Tr}\,(\hat \rho _{\{\hat G\}}\hat G_\nu )&=&\overline{G}_\nu \,,
\,\,\,\,\nu = 1,2,...,n \, ,
\label{2.8}
\ee
that is,  the the normalization condition and the constraints imposed by the
results of the measurement.
To estimate the unknown density operator in the most
reliable way we utilize the Jaynes principle of maximum entropy
({\em MaxEnt} principle) 
\cite{Jaynes1957,Fick1990,Kapur1992,Katz1967}
according to which among those operators which fulfill the
constraints (\ref{2.8}) the most reliable 
reconstruction (estimation)
$\hat{\rho}_{r}$ is the one 
with the maximal value of the von Neumann entropy
$S(\hat{\rho})=-{\rm Tr}(\hat{\rho}\ln\hat{\rho})$:
\be
S(\hat{\rho}_{r})=
{\rm max}\left[S(\hat{\rho}_{{\{\hat G\}}}); \forall \hat{\rho}_{{\{\hat
G\}}}\right].
\label{2.8a}
\ee

As shown by E.Jaynes \cite{Jaynes1957} 
The operator which fulfills the constraints (\ref{2.8}) and simultaneously
maximizes the von Neumann entropy can be expressed in the generalized
canonical form
\be
{\hat \rho _{r}={1 \over {Z_{\{\hat G\}}}}
\exp \,(-\sum\limits_\nu  {\lambda _\nu \hat G_\nu })};
\label{2.11}
\ee
where
\be
{Z_{\{\hat G\}}(\lambda _1,...,\lambda _n)=
{\rm Tr}[\exp (-\sum\limits_\nu  {\lambda _\nu \hat G_\nu })]},
\label{2.12}
\ee
is the generalized partition function and 
$\lambda_\nu$ are the Lagrange multipliers.
The Lagrange multipliers $\lambda_\nu$ are chosen so
that the density operator (\ref{2.11}) fulfills the constraints (\ref{2.8})
imposed by the results of the measurement.
It is then obvious that the meanvalues $\overline{G}_\nu$ 
which determine the density operator 
are related to the Lagrange multipliers via 
the derivatives of the partition function 
\be
\overline{G}_\nu ={\rm Tr}(\hat \rho_{r}\hat G_\nu )=
-{\partial  \over {\partial \lambda_\nu }}
\ln Z_{\{\hat G\}}(\lambda _1,...,\lambda _n).
\label{2.13}
\ee
If we solve the last equation with respect to 
the Lagrange multipliers we can express them in terms of the measured
meanvalues 
\be
\lambda _\nu =\lambda _\nu (\overline{G}_1,...,\overline{G}_n) .
\label{2.15}
\ee
When we substitute the Lagrange multipliers (\ref{2.15}) into the
expression for the generalized canonical density operator
(\ref{2.11}) we obtain the explicit expression for the 
reconstructed (estimated)  density operator.

As a typical example of the application of the MaxEnt principle we can
consider a measurement of a single-mode electromagnetic field, modeled
as a harmonic oscillator. Imagine, that as a result of the measurement
we know the mean photon number $\bar{n}$ in the given field mode.
Certainly, there are (infinitely) many quantum states of a single-mode 
electromagnetic field (e.g., a Fock state, a coherent state, a squeezed
state, etc.) with the given mean photon number. So the question is:
Which is the best (most reliable) estimation of the measured state
given the mean photon number is known? The mean photon number is
in some sense
the least available information about the measured state. Consequently,
the state is least determined. On the other, 
pure states are completely determined which is reflected by the fact
that they 
have a zero von Neumann entropy.
Therefore we expect that the most reliable 
 reconstruction in the given case is a statistical mixture which is
determined just by a single parameter - the mean photon number.
It is well known that a statistical mixture which is parametrized
just by a single parameter is a thermal state. This statistical mixture
of Fock states is characterized by a temperature, or the corresponding 
mean photon number. In addition, for a given temperature (mean photon
number) the thermal state exhibits the largest von Neumann entropy.
Consequently, from the Jaynes principle of the maximum entropy it follows
that if from a measurement only a mean photon number is known that the
most reliable estimation of the measured state is the thermal state.

 The MaxEnt principle is not the only criterion 
how to choose an appropriate density operator among those 
$\hat \rho _{\{\hat G\}}$ which fulfill the constraints (\ref{2.8}).
Based on an intuition or some addition {\it a priori} knowledge
one can apply other criteria.  
For example,  the {\it maximum likelihood} principle has been 
adopted successfully for estimation of quantum states \cite{Hradil2000}.
Although this reconstruction scheme can result in  nonphysical 
estimations (e.g., density operators which are not properly normalized,
etc).
In general, a consistent reconstruction scheme has to avoid 
nonphysical results (e.g. occurrence of negative probabilities). 
In the MaxEnt reconstruction a physical estimate is guaranteed by 
the canonical form of the density operator (\ref{2.11}).  
The MaxEnt principle is {\em the most conservative
assignment in the sense that it does not permit one to draw any conclusions
not warranted by the data}. From this point of view the MaxEnt principle
has a very close relation (or can be understood as the generalization) of
the Laplace's principle of {\em indifference}  which
states that where nothing is known one should choose a constant valued
function
to reflect this ignorance. Then it is just a question how to quantify a
degree
of this ignorance. If we choose an entropy to quantify the ignorance, then
the relation between the Laplace's indifference principle and the Jaynes
principle of the Maximum Entropy is transparent, i.e. for a constant-valued
probability distribution the entropy takes its maximum value.

The MaxEnt reconstruction has been applied for various quantum 
systems, such as light field mode, spin systems \cite{Buzek1996}. 
In what follows we adopt it for the reconstruction 
of vibrational states of neutral atoms. 
We assume the experimental setup  realized 
by the group of C. Salomon at the ENS in Paris 
\cite{Bouchoule99,Morinaga99}.

\section{Reconstruction of motional states of neutral atoms}
\label{sec3}

Recently, experimental manipulations of motional quantum 
states of neutral atoms have been reported by the group of C. Salomon 
in Paris \cite{Bouchoule99,Morinaga99}. 
Cold Cs atoms can be cooled into specific quantum states of a far detuned 
1D optical lattice. The optical lattice is induced by the interference 
of two laser beams. Along the vertical $z$ axis a periodic potential 
of ``harmonic'' micro-wells is produced with a period of 665 nm and 
with an amplitude of about 0.2 $\mu$K \cite{Morinaga99}. 
The vertical oscillation frequency in a micro-well at the center of the trap 
is $\omega_z/2\pi=85$ kHz. The corresponding ground state has the rms size 
$\Delta z_0=\sqrt{\hbar/2m \omega_z}\approx 21$ nm and 
$\Delta p_0/m=\sqrt{\hbar\omega_z/2m}\approx 11$ mm/s 
is its rms velocity width.
The trapped cloud of neutral Cs atoms has a nearly Gaussian shape
with a vertical rms size $\Delta \xi_0=53~\mu$m. 
With the help of deterministic manipulations the neutral atoms can be 
prepared in non-classical  1D motional states 
along the vertical axis such as squeezed states, number states, or specific 
superpositions of number states \cite{Morinaga99}.
The measurement of the prepared quantum state $\hat\rho$ is performed 
as follows: The system is evolved within 
the harmonic potential during the time $\tau$. Then the lasers are
turned off and the system undergoes the ballistic expansion (BE).
After the time of flight $T=8.7$ ms a 2D absorption image of the cloud 
is taken in 50 $\mu$s with a horizontal beam \cite{Morinaga99}.
 Integration of 2D absorption images in the horizontal direction 
gives us the spatial distribution along the vertical $z$ axis. 
Therefore we will consider only 1D quantum-mechanical system 
along the vertical axis.

To confirm that a desired quantum state has been obtained (engineered)
one can compare  the spatial distributions along the vertical axis 
with the predicted ones.
The coincidence of these  spatial distributions 
is a necessary but not the sufficient requirement.
A complete verification of the fidelity of the preparation 
of desired  quantum states requires a 
quantum state reconstruction procedure. 
In order to perform this task we adopt 
the MaxEnt principle \cite{Buzek1996}. 
To do so we utilize 
a close analogy between  quantum homodyne tomography \cite{Leonhardt1997} 
and the BE absorption imaging 
for the case of the point-like cloud (with the rms size equal to zero).

\subsection{Quantum tomography via MaxEnt principle}
\label{sec6}
Quantum tomography is based on the inverse Radon transformation
of the the measured 
probability density distributions 
$w_{\hat{\rho}}(x_{\theta})$ 
for rotated quadratures
$\hat{x}_{\theta}=\frac{1}{\sqrt{2}}\left(\hat{a}{\rm e}^{-i\theta}
+\hat{a}^{\dagger}{\rm e}^{i\theta}\right)$ 
\cite{Vogel1989,Leonhardt1997}.
These distributions 
can be represented as a result of the measurement
of the continuous set of projectors 
$|x_{\theta}\rangle\langle x_{\theta}|$. 
Based on the measurement of the distributions 
$w_{\hat{\rho}}(x_{\theta})$ for all values of 
$\theta\in [0,\pi]$
we can formally 
reconstruct the density operator according to the
formula \cite{Buzek00}
\be
\hat{\rho}_{r}=\frac{1}{Z_{0}}
\exp\left[-\int_{0}^{\pi} d\theta\,
\int_{-\infty}^{\infty}dx_{\theta}\,
|x_{\theta}\rangle\langle x_{\theta}| \lambda(x_{\theta})\right],
\label{6.3}
\ee
where the Lagrange multipliers $\lambda(x_{\theta})$
are given by an infinite set of equations
\be
w_{\hat{\rho}}(x_{\theta})=\sqrt{2\pi}
\langle x_{\theta}| \hat{\rho}_{r}|x_{\theta}\rangle;
\qquad \forall x_{\theta} \in (-\infty,\infty).
\label{6.2}
\ee
If the distributions 
$w_{\hat{\rho}}(x_{\theta})$ are measured for all values of $x_\theta$ and
all angles  $\theta$ then the density operator $\hat{\rho}_{r}$
is reconstructed precisely and is equal to density operator
obtained with the help of 
the inverse Radon transformation or with the help of the
pattern functions (for more details see  \cite{Buzek00}).

In practical experimental situation (e.g., see 
the experiments by Raymer et al. \cite{Smithey93}
and by Mlynek et al. \cite{Kurtsiefer96}) it is impossible to measure the
distributions 
$w_{\hat{\rho}}(x_{\theta})$  for all values of $x_\theta$ and
all angles  $\theta$. What is measured are distributions (histograms)
for finite number $N_\theta$ quadrature angles 
 $\theta$ and the finite number $N_x$ 
of ``bins'' for quadrature operators. This means that 
practical  experiments
are associated with an observation level specified by a {\em finite} number
of observables 
\be
\hat{F}_{jk}=|x_{\theta_k}^{(j)}\rangle\langle x_{\theta_k}^{(j)}|
\ee
with the number of quadrature angles equal to $N_\theta$ and the
number of bins for each quadrature equal to $N_x$.
We can therefore assume that from the measurement 
of the observables $\hat{F}_{jk}$ the meanvalues
$\overline{F}_{jk}$  are determined 
(these meanvalues correspond to ``discretized'' quadrature distributions).
In addition it is usually the case that the mean excitation  number
of the state is known (measured) as well.

The operators $\hat{F}_{jk}$ together with $\hat{n}$ 
form a specific observation level
corresponding to the {\em incomplete tomographic measurement}. 
In this case we can express the generalized 
canonical density  operator in the form  
\begin{equation}
\hat{\rho}_{r} = \frac{1}{Z} \exp\left(-\lambda_n \hat{n} - 
\sum_{j=1}^{N_x}
\sum_{k=1}^{N_{\theta}}
\lambda_{j,k}
\hat{F}_{jk} 
\right)
\label{6.4}
\end{equation}
The knowledge of the mean photon number is essential
for the {\em MaxEnt} reconstruction because it formally regularizes the
{\em MaxEnt} reconstruction scheme  (the generalized partition function is
finite in this case). 

\subsection{Motional states of atoms via MaxEnt principle: Formalism}

In the quantum homodyne tomography the probability distributions 
are measured for the rotated quadrature operators 
$\hat{x}_{\theta}$. The annihilation and creation operators
of motional quanta, $\hat{a}$ and $\hat{a}^\dagger$,
are related to the position and momentum operators, 
$\hat{z}$ and $\hat{p}$, via expressions 
$\hat{z}=\frac{1}{\sqrt{2}} (\hat{a}+\hat{a}^\dagger)$
and $\hat{p}=\frac{1}{\sqrt{2}} i (\hat{a}-\hat{a}^\dagger)$, respectively.
The angle $\theta$ of the quadrature operator corresponds 
to $\omega_z \tau$
and vertical ``cuts'' of the absorption images (taken after the BE)
can be associated with quadrature probability distributions.
However, for a real physical situation with a nonzero rms size of 
the cloud the vertical ``cuts'' of absorption images correspond to 
a coarse-grained  quadrature probability distributions. 
In particular, the vertical cuts of measured absorption images
(taken in 2D) give us (after integration along the horizontal direction)
the spatial distribution along the vertical axis. The spatial 
distribution along the vertical $z$ axis can be expressed as 
\be
\overline{F}_\tau(z)=T^{-1} \int F_0(\xi_0) P_\tau((z-\xi_0)/T) d\xi_0 \, ,
\ee
where $F_0(\xi_0)$ is the initial spatial distribution of the cloud
in the $z$-direction (i.e., a Gaussian distribution 
with the rms size $\Delta \xi_0$).
The function
$P_\tau(v)$ denotes the velocity probability distribution of the measured 
quantum state which has been evolved for time $\tau$ in the harmonic
potential before the BE, i.e.
\be
P_\tau (v)=\vert \langle v|\psi(\tau) \rangle \vert^2,
\qquad \vert\psi(\tau)\rangle=\hat{U}(\tau) \vert \psi(0) \rangle 
.\ee
Here $\hat{U}(\tau)=\exp(-i\hat{H}\tau/\hbar)$ represents the 
time-evolution operator for the harmonic oscillator with the Hamiltonian
$\hat{H}=\hat{p}^2/2m +m\omega_z^2 \hat{z}^2/2$.   
Now we can treat the measured ``cuts'' as 
the mean values of specific observables:
$\overline{F}_\tau(z)=\mbox{Tr~} [\hat\rho \hat{F}_\tau(z)]$.
In practice just a few discrete times $\tau_j$ ($j=1,\ldots,N_\tau$) are 
considered 
and the $z$ coordinate is discretized into 
the bins $z_k$ ($k=-N_z,\ldots,N_z$) 
of a given resolution $\Delta z$. 
The set of operators which enters
the equation (\ref{6.4}) for 
 the MaxEnt reconstruction 
then takes  the form
\be
\hat{F}_{\tau_j} (z_k) =T^{-1} \int F_0(\xi_0) \hat{U}^\dagger(\tau_j) 
\bigg|\frac{z_k-\xi_0}{T}\bigg\rangle \bigg\langle \frac{z_k-\xi_0}{T}\bigg| 
\hat{U}(\tau_j) d\xi_0
\qquad (j=1,\ldots,N_\tau; k=-N_z,\ldots,N_z)
.\ee
We have already commented that the 
operator of mean phonon number $\hat{n}$ 
is added to the set of observables $\{ \hat{F}_{\tau_j}(z_k) \}$.
Knowledge of the mean excitation number $\bar{n}$ is  essential 
in the case of an incomplete set  of observables \cite{Buzek1996}. 
Knowledge of the mean excitation number leads to a natural 
``truncation'' of the Hilbert space. 
The inclusion of the mean phonon number into the MaxEnt reconstruction 
scheme does not represent its limitation as the mean energy 
represents one of basic characteristics of any system 
which should be inferred from the measurement.

The experimental ``cuts''  of the BE absorption images 
[$\hat{F}_{\tau_j} (z)$] can be taken at few selected times, for example 
$\omega_z \tau_j=0,\pi/4,\pi/2, 3\pi/4$ ($N_\tau=4$).
To perform the reconstruction we have to determine the Lagrange
multipliers $\{ \lambda_{j,k} \}$ and $\lambda_n$ associated with 
$\{ \hat{F}_{\tau_j} (z_k) \}$ and $\hat{n}$, respectively,
in the expression for the generalized canonical density operator
(\ref{6.4}).
The Lagrange multipliers can be determined via the minimization of 
a deviation function $\Delta F$  with respect to the measured data, i.e. 
\be
\Delta F &=& \sum_{j,k} w_{j,k} \left\{ 
\overline{F}_{\tau_j}(z_k) - {\rm Tr}\left( \hat{\rho}_r \hat{F}_{\tau_j}(z_k) 
\right) \right\}^2 
\nonumber
\\
&+& w_{\bar{n}}
\left\{ \bar{n} - {\rm Tr} \left(\hat{\rho}_r \hat{n} \right) \right\}^2 
.\ee
Here $\{ w_{j,k} \}$ and 
$w_{\bar{n}}$ represent positive weight factors 
for particular observables. Without any  prior knowledge about the state 
we can take 
for simplicity $w_{i,j}=1$. The weight factor $w_{\bar{n}}$ associated 
with the mean phonon number can be chosen according to our preference
either to fit better the ``cuts'' of the BE images or the mean phonon number.
In the case of the perfect measurement and the complete reconstruction 
the result has to be independent of the choice of the weight factors
(in this case we can take $w_{\bar n}=1$). 
The weight factors could be also associated 
with the {\em prior} information about 
the dispersion of the measured observables. 
In particular, the weight factors can be taken 
as $w_\nu\sim\sigma_\nu^{-2}$ to reflect the knowledge of variances 
$\sigma_\nu$ for the measured observables $\hat{G}_\nu$. 
When the mean values of the observables for the MaxEnt estimate 
$\hat{\rho}_r$ fit within desired interval $\overline{G}_\nu\pm \sigma_\nu$ 
then contributions of the observables to the deviation function 
$\Delta F$ are of the same order ($\sim 1$). However, 
in our case we do not assume the knowledge of variances for the 
measured discretized probability distributions (taking $w_\nu=1$).

Once the Lagrange multipliers are numerically fitted, the result of the
reconstruction -- the generalized canonical density operator $\hat\rho_r$ -- 
can be visualized, for example, via the corresponding Wigner function 
\cite{Wigner} which 
can be defined as a particular Fourier transform of the
density operator $\hat{\rho}$ of a harmonic oscillator
expressed in  the basis of the eigenvectors $|q\rangle$ of the
position
operator $\hat{q}$:
\be
W_{\hat{\rho}}(q,p)\equiv
\int_{-\infty}^{\infty} d\zeta \langle q-\zeta/2|\hat{\rho}|q+
\zeta/2\rangle {\rm e}^{ip\zeta}.
\label{3.16}
\ee

\subsection{Numerical simulation}

To test our reconstruction procedure let us  consider 
the reconstruction of the Wigner function of the motional quantum 
state $\vert \psi(0)\rangle=(|0\rangle+|1\rangle)/\sqrt{2}$ 
of  Cs atoms trapped in the optical lattice.
This kind of states has been demonstrated in recent experiments
\cite{Morinaga99}.
We assume the following setup parameters: $\omega_z/2\pi=80$ kHz, 
the rms size of the ground state $\Delta z_0=22$ nm, 
the rms velocity width $\Delta p_0/m=11$ mm/s
and the rms width of the cloud of the atoms about 60 $\mu$m.
Before BE (with BE time $T=8.7$ ms) the atoms evolve within 
the harmonic trapping potential for $\tau=0, 1.6, 3.2, 4.8$ $\mu$s. 
As the input for the reconstruction via the MaxEnt principle four 
vertical ``ideal'' cuts of the BE absorption images 
are taken as shown in insets of Fig.~1.
In addition, for the phonon number operator $\hat{n}$ which
is included in the set of measured observables (see discussion above)
we assume the mean value $\bar{n}=0.5$.
The result of the ideal reconstruction is shown in Fig.~1.
The fidelity of the measured and the reconstructed quantum states 
is close to unity which means a perfect reconstruction with 
$\Delta F=10^{-10}$, entropy $S=10^{-7}$, $\Delta \rho=10^{-8}$
has been achieved. 
 Here $\Delta\rho= \sum_{m,n} |(\hat\rho-\hat\rho_r)_{mn}|^2$
denotes a deviation of the original and reconstructed density operators.

\begin{figure}
\begin{minipage}{10cm}
\includegraphics[height=8cm,clip]{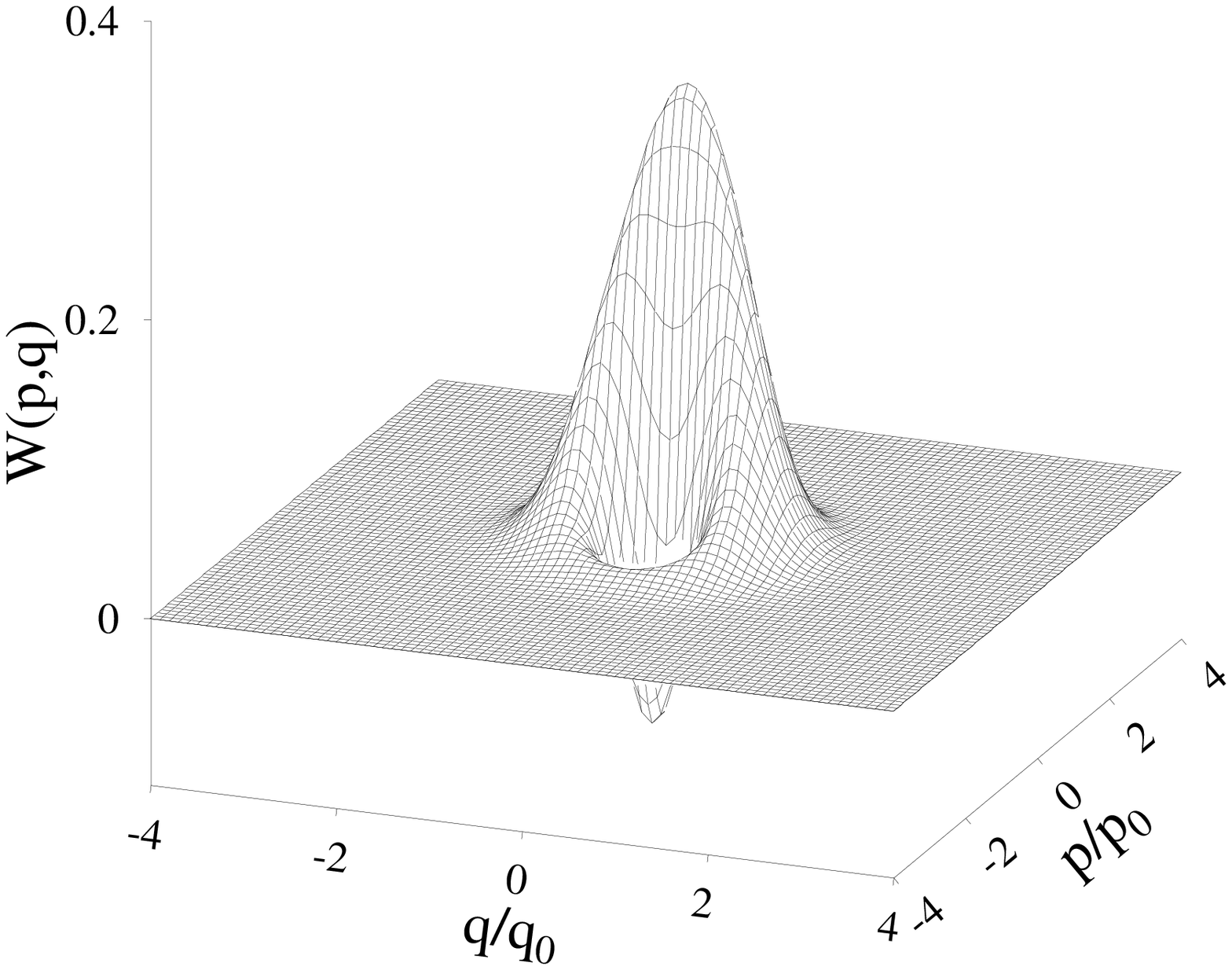}
\end{minipage}
\begin{minipage}[b]{6cm}
\includegraphics[height=5cm,clip]{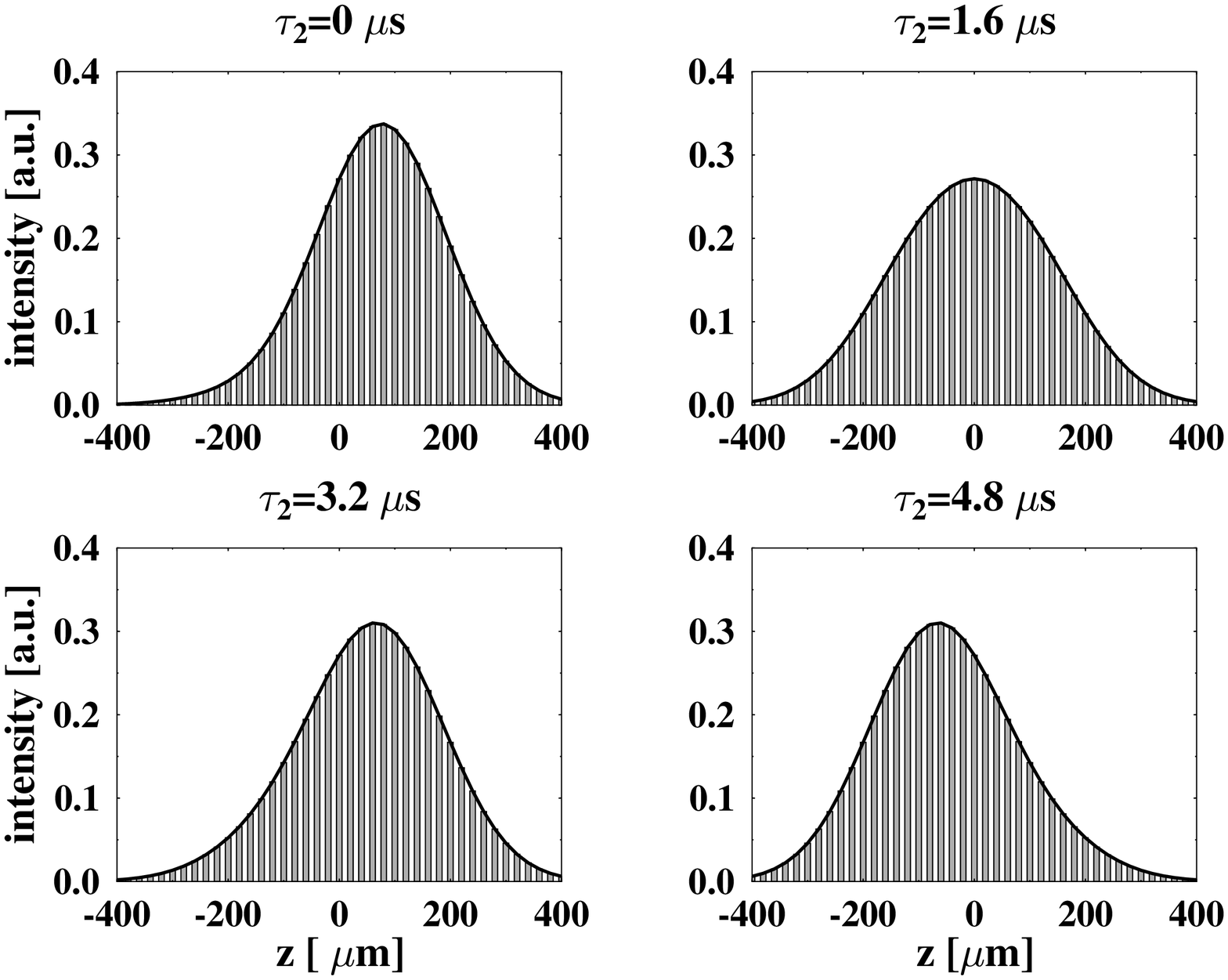}
\end{minipage}
\caption{
Numerical simulation of the reconstruction of the 
Wigner function of the motional quantum 
state $(|0\rangle+|1\rangle)/\sqrt{2}$ of  Cs atoms trapped 
in the optical lattice (assuming $\omega_z/2\pi=80$ kHz, the rms size
of the ground state $\Delta z_0=22$ nm and the rms velocity width 
$\Delta p_0/m=11$ mm/s). 
As shown in insets 
for the reconstruction via the MaxEnt principle four 
vertical cuts of the absorption images (with BE time $T=8.7$ ms)
have been taken. The histograms in the insets correspond to the measured
data while the solid lines are obtained from the reconstructed Wigner
function (i.e. they correspond to reconstructed marginal distrubutions).
Before BE the atoms evolve within the trapping 
potential for the times $\tau=0, 1.6, 3.2, 4.8$ $\mu$s.  
In addition, the mean number of motional quanta
$\bar{n} =0.5$ and the rms width of the cloud of the atoms 
about 60 $\mu$m have been assumed.}
\end{figure}

Obviously, in a real measurement the measured values are always 
fluctuating around the exact ones due to an experimental noise. 
Therefore we simulate a non-ideal measurement introducing  random 
fluctuations to the measured values of observables. 
It means that instead of the ideal values $\overline{F}_{\tau_j}(z_k)$
we use for the MaxEnt reconstruction procedure the fluctuating 
(``noisy'') values 
\be
\overline{F}^\prime_{\tau_j}(z_k)=\overline{F}_{\tau_j}(z_k) +
\eta \xi_{j,k} \left(\overline{F}_{\tau_j}(z_k)\right)^{1/2} 
\label{3.17}
.\ee
Here $\eta$ is a relative-error parameter which characterizes 
the quality of the measurement and $\{ \xi_{j,k} \}$ represents  
a Gaussian noise for observables. 
The result of the reconstruction is shown in Fig.~2 for $\eta=0.1$. 
Noisy mean values of the observables are shown in the insets.
Despite of a significant relative error the reconstruction is
almost perfect with the fidelity of the measured and the reconstructed
states still close to one 
($\Delta F=0.16$, entropy $S=0.01$, $\Delta\rho=0.05$).
The minimum value of the deviation function 
$\Delta F=0.16$ can serve also as a measure
of the imperfection of the given measurement (due to  a technical noise)
\cite{foot1}.

\begin{figure}
\begin{minipage}{10cm}
\includegraphics[height=8cm,clip]{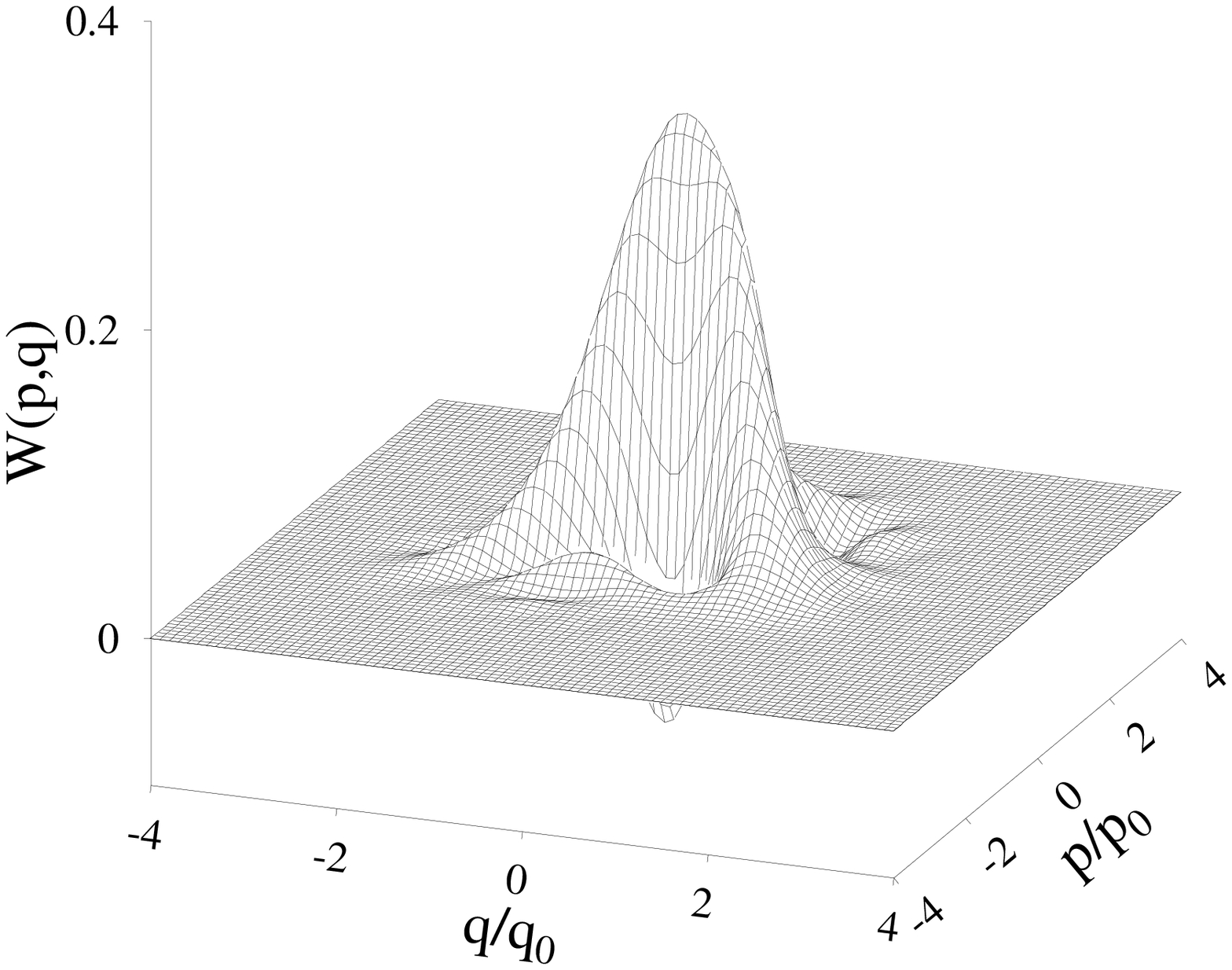}
\end{minipage}
\begin{minipage}[b]{6cm}
\includegraphics[height=5cm,clip]{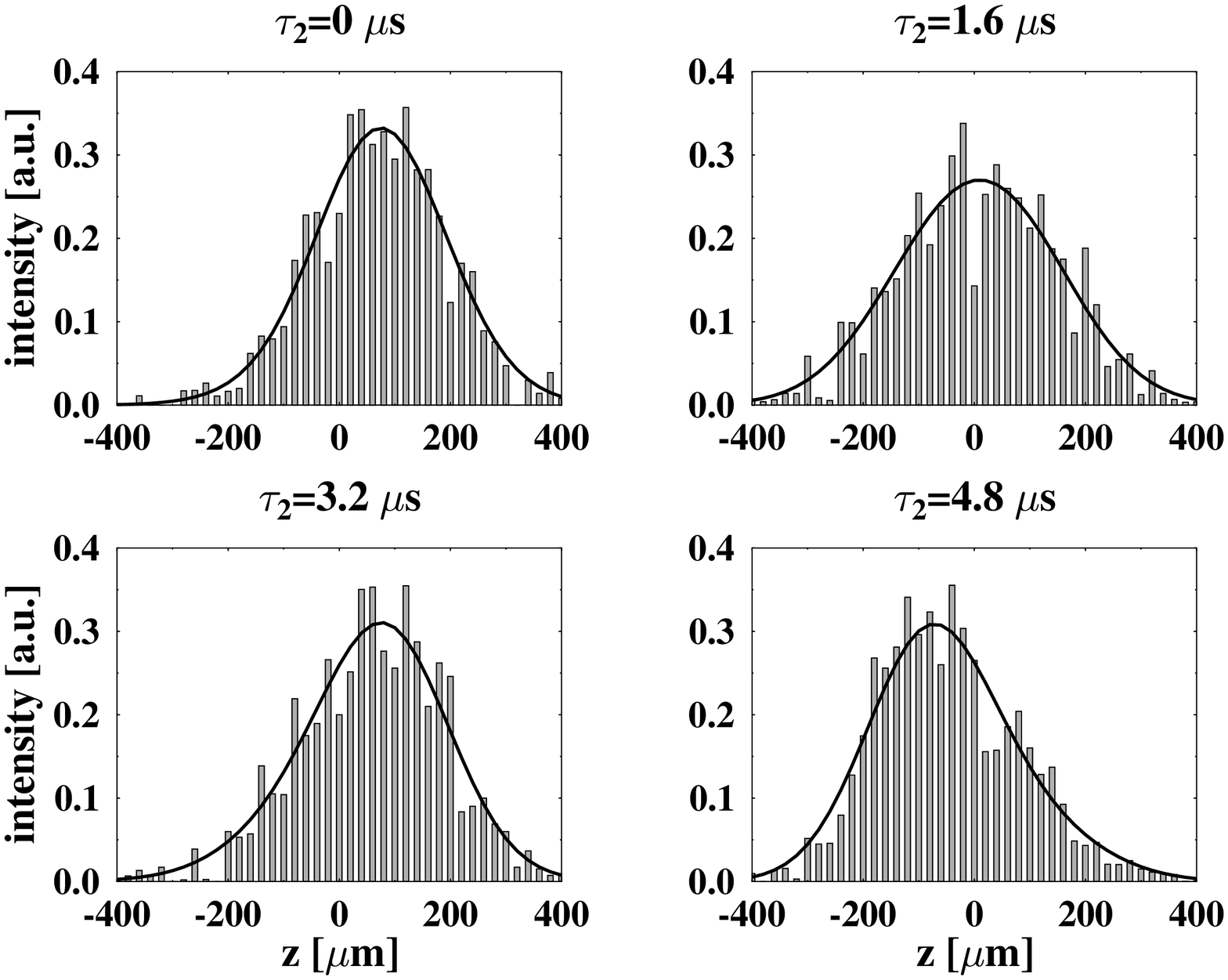}
\end{minipage}
\caption{
Numerical simulation of the reconstruction of the 
Wigner function of the atomic motional quantum 
state $(|0\rangle+|1\rangle)/\sqrt{2}$ for the same settings 
as in  Fig.~1.
Four vertical cuts of the absorption images taken for reconstruction
(see the insets) are fluctuating randomly around their ideal values 
shown in Fig.~1 with the relative error $\eta=0.1$.
In addition, the mean phonon number
$\bar{n}^\prime =0.6$ has been considered.}
\end{figure}

A typical non-classical state which we can utilize  for a further test 
is the even coherent state 
${\mathcal N}_e (|\alpha\rangle+|-\alpha\rangle)$
which is a superposition of two coherent states with opposite phases
\cite{Buzek1995}.
For the amplitude $\alpha=\sqrt{2}$ we obtained 
$\Delta F=10^{-8}$, the entropy $S=0.026$ and $\Delta \rho=10^{-4}$ 
(under assumption that the exact mean phonon number 
$\bar{n}=1.928$ is known).
In the case of the imperfect measurement with $\eta=0.1$
the reconstruction leads to $\Delta F=0.14$, entropy $S=0.13$, 
and $\Delta\rho=0.06$ for $\bar{n}=2.09$. 
The fidelity of the reconstructed and the measured states is  
in this case also close to one.

In order to model a technical noise in the measurement
we have been considered
 Gaussian fluctuations proportional to the square root of 
the mean values.
It means that tails of the ``cuts'' of BE images do not introduce
a significant error (compare insets in Fig.~1 and Fig.~2).
However, in the current measurements the situation seems to be different 
and the fluctuations do not decrease with the amplitude 
of the expected values.

The fundamental question in the context of the
{\em MaxEnt} reconstruction of states from incomplete
tomographic data 
is whether the quality of the reconstruction 
can be improved using additional data from subsequent time moments $\tau$
and how many such time moments $\tau$ are required for the complete 
reconstruction of the unknown state $\hat{\rho}$.
We have shown recently \cite{Buzek00} that for the 
quantum tomography just {\em three} quadrature distribution 
are sufficient for a complete reconstruction using the MaxEnt principle 
(in the case of the perfect measurement).
This corresponds to the ideal case 
without the spatial dispersion of the cloud of atoms,
i.e. the choice with $\omega_z \tau_j=0,\pi/4,\pi/2$
($N_\tau=3$) is sufficient for $\Delta \xi_0\to 0$.  
Obviously, in experiments with neutral atoms 
the spatial size of the atomic cloud is nonzero.
However, in the case of the ideal measurement 
three BE absorption images 
 associated with three ``rotations'' $\omega_z \tau_j$ 
are  still sufficient 
for a complete reconstruction of tested examples of quantum states.
On the other hand, it seems that for higher mean phonon numbers 
 the spatial distributions along the vertical axis 
which are directly determined from absorption images 
should be known with improving precision 
(and on a wider interval of values  as well).  
In the above  examples we have considered for convenience 
 BE images for four ``rotations'' 
($N_\tau=4$) which results in a very good reconstruction.

\begin{figure}
\begin{minipage}{10cm}
\includegraphics[height=8cm,clip]{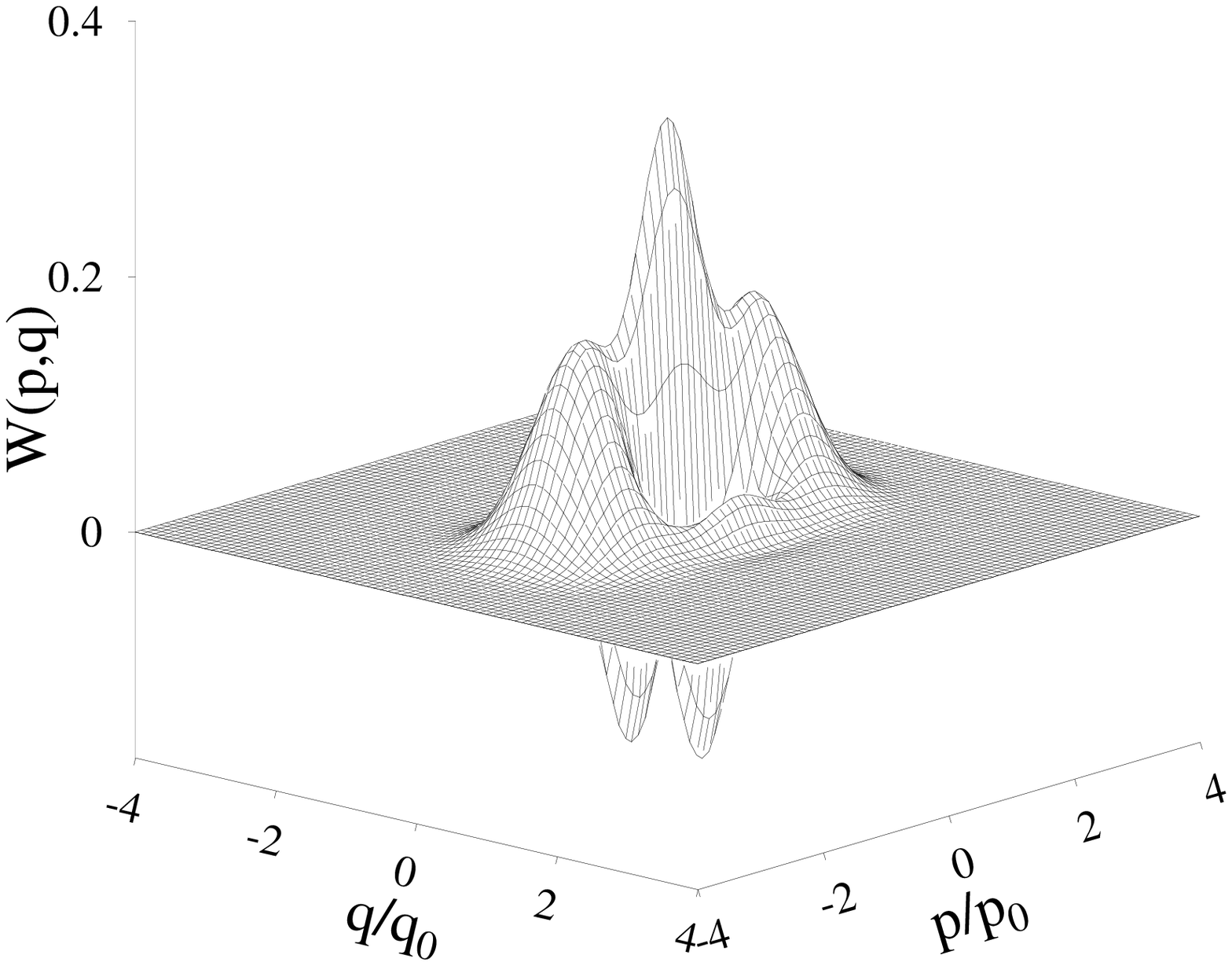}
\end{minipage}
\begin{minipage}[b]{6cm}
\includegraphics[height=5cm,clip]{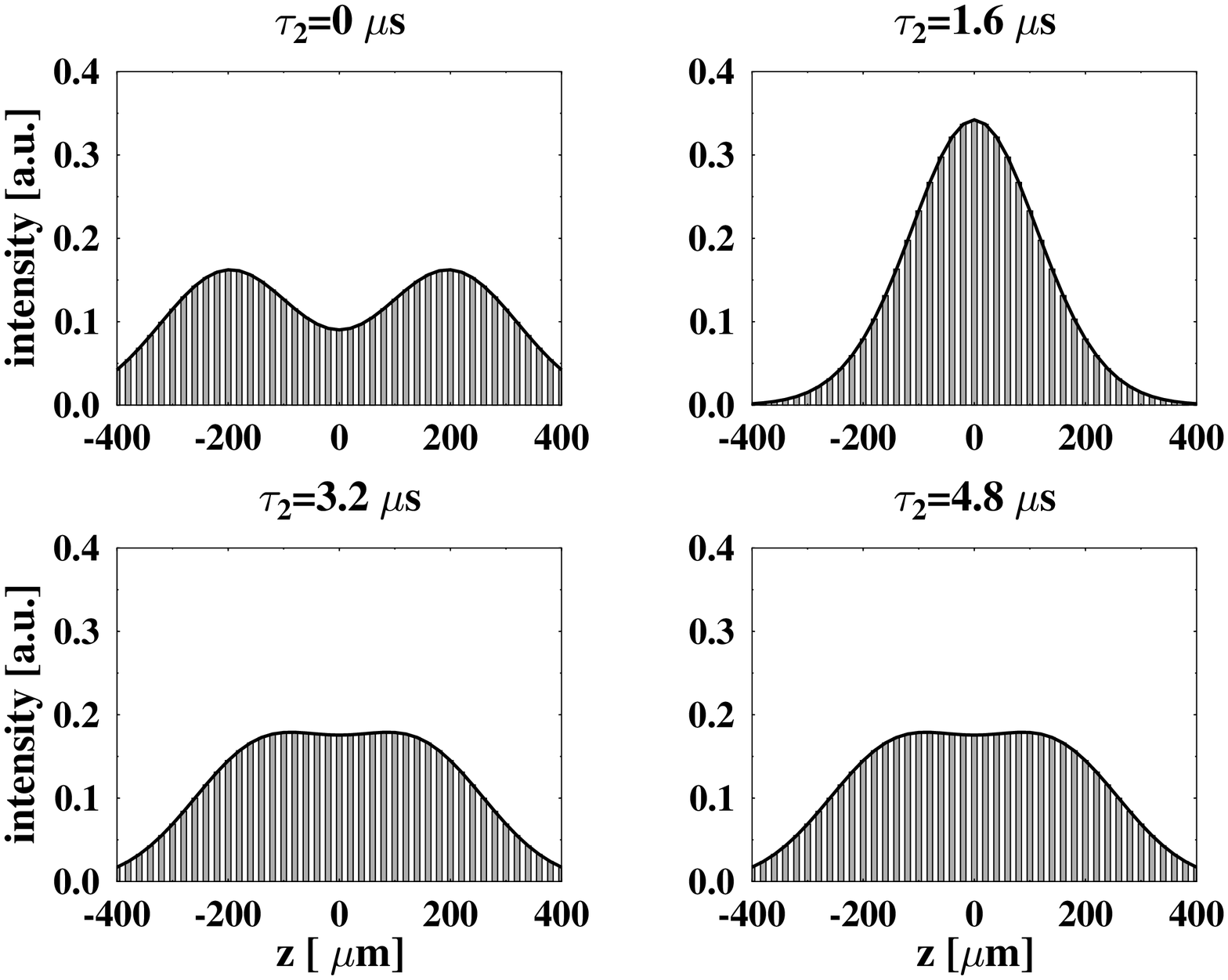}
\end{minipage}
\caption{
Numerical simulation of the reconstruction of the 
Wigner function of the motional quantum 
state ${\cal N}_e (|\alpha\rangle+|-\alpha\rangle)$
with $\alpha=\sqrt{2}$ for the case of the ideal measurement.
The mean number of motional quanta $\bar{n}=1.928$.
Other settings are same as in Fig.~1.}
\end{figure}

\begin{figure}
\begin{minipage}{10cm}
\includegraphics[height=8cm,clip]{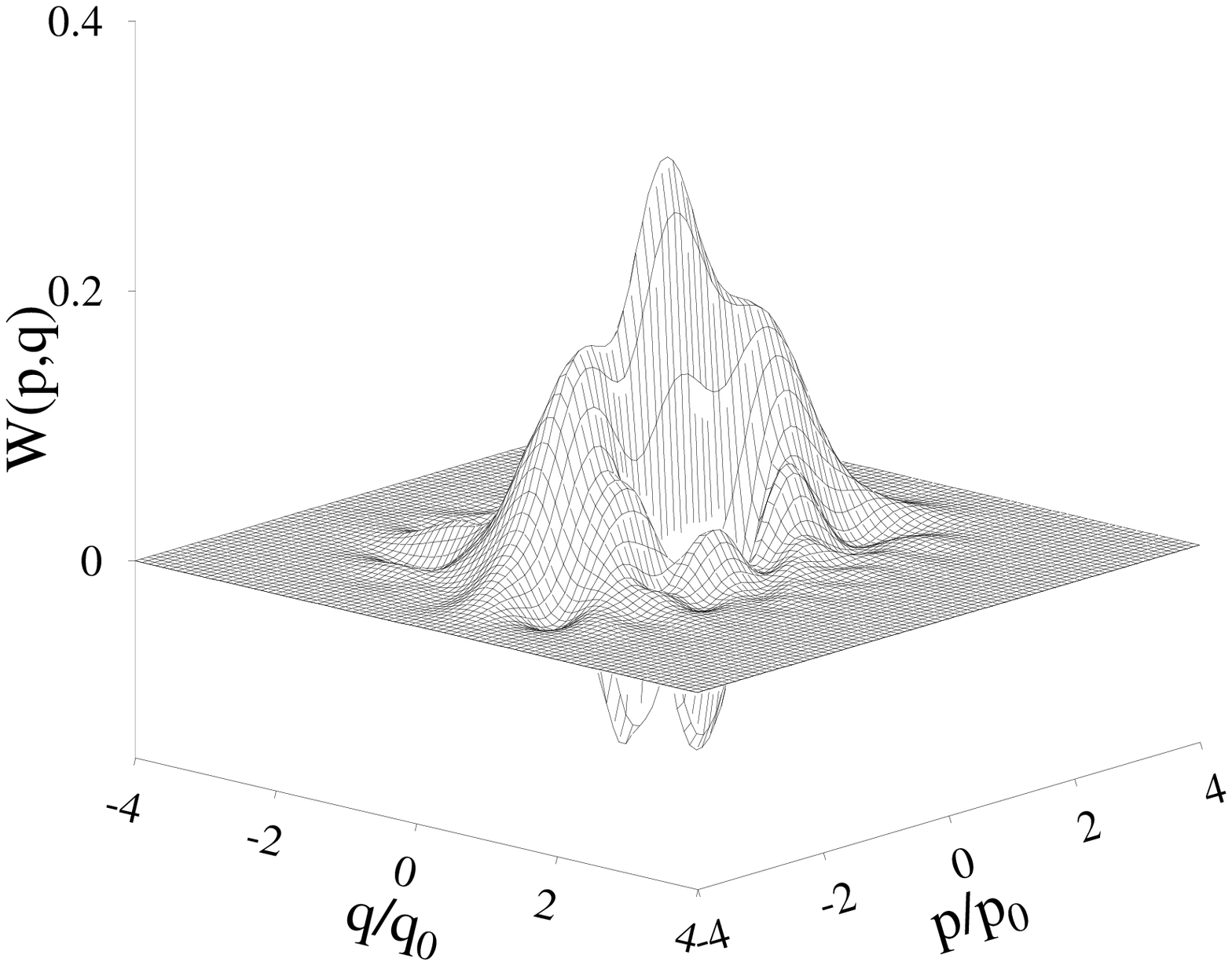}
\end{minipage}
\begin{minipage}[b]{6cm}
\includegraphics[height=5cm,clip]{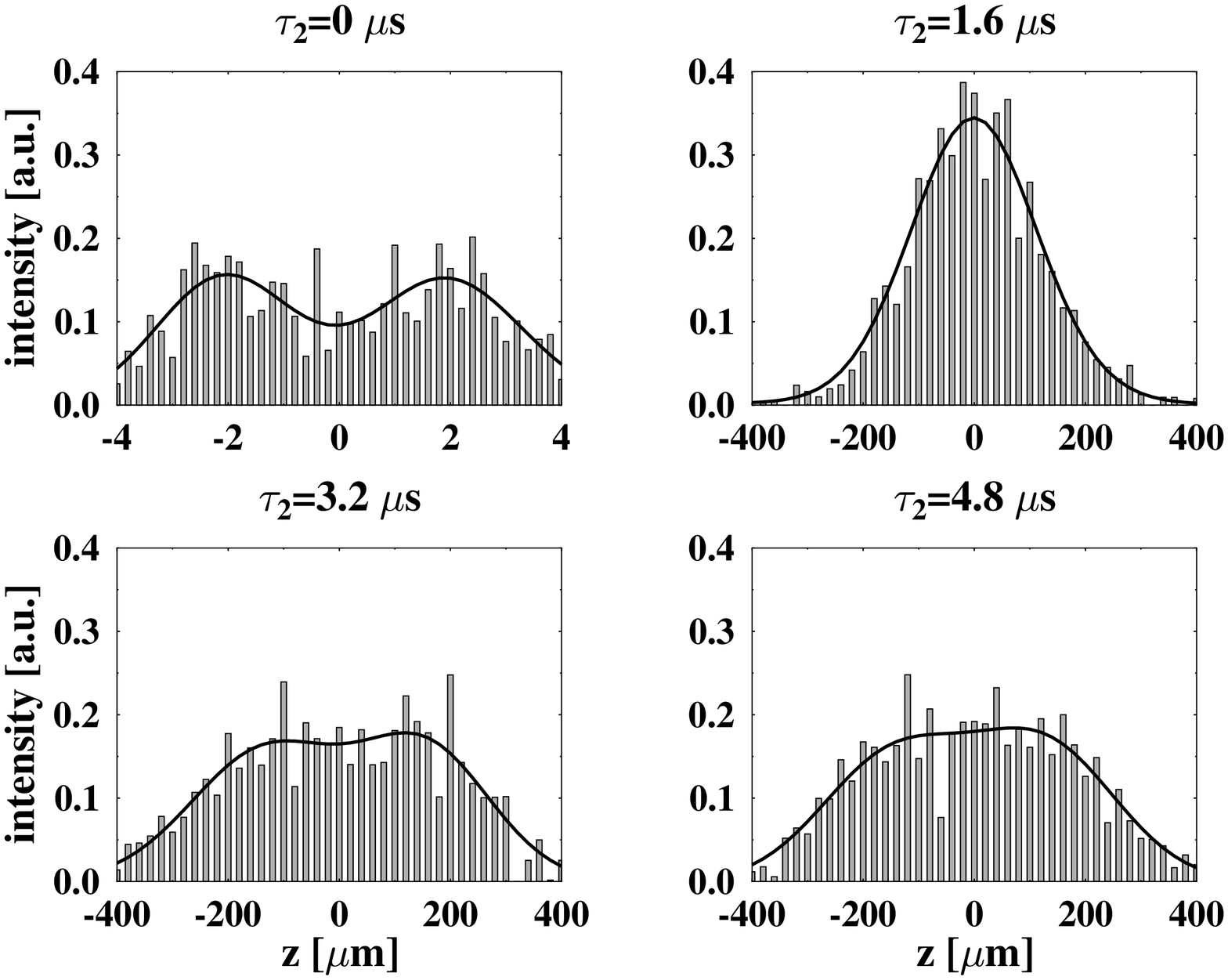}
\end{minipage}
\caption{
Numerical simulation of the reconstruction of the 
Wigner function of the motional quantum 
state ${\cal N}_e (|\alpha\rangle+|-\alpha\rangle)$
with $\alpha=\sqrt{2}$ for the case of noisy measurement with $\eta=0.1$.
The ``measured'' mean number of the motional quanta 
$\bar{n}^\prime =2.09$. Other settings are the same as in Fig.~1.}
\end{figure}

\section{Reconstruction from experimental data}
\label{sec4}

In what follows we will apply the MaxEnt reconstruction scheme
to the data obtained at the ENS in Paris
\cite{Salomon00}. 
Firstly we note that the unknown quantum state should belong to a Hilbert 
subspace which can be determined easily. 
Thus we can limit ourselves to the subspace spanned by Fock (number) 
states $|0\rangle,|1\rangle,\ldots |N-1\rangle$. 
The upper bound on the accessible phonon number $N$ is given by experimental 
limitations such as, for example, a feasible depth of micro-wells of 
the optical lattice and the validity of the harmonic potential approximation.
For recent experiments $N$ has been typically of the order of 10.
This value is large 
enough to demonstrate the preparation of many non-classical
states but on the other hand
excludes  highly squeezed states from a coherent
processing.

Let us consider the experimental arrangement used in Paris \cite{Salomon00}
with the parameters: $\omega_z/2\pi=80$ kHz, 
the rms size of the ground state $\Delta z_0=22$ nm, 
the rms velocity width $\Delta p_0/m=11$ mm/s,
the rms width of the cloud of the atoms about 60 $\mu$m
and BE time $T=8.7$ ms. 
Initially the atoms are prepared in a well defined motional state
$|\psi_0\rangle$ (e.g. in the vacuum state $|0\rangle$).
Then the optical lattice is switched off for the time period $t_1$ 
during which the atoms evolve freely towards the state 
$|\psi_1\rangle=\exp(-i t_1 \hat{p}^2/2m) |\psi_0\rangle$. 
Next, the optical lattice is again switched on for the time $\tau$ during 
which the atoms evolve within the harmonic trapping potential. 
The measurement is performed after the BE time.
The first two stages can be considered as the preparation of the 
state $|\psi_1\rangle$. 
 After its ``rotation'' by $\omega_z \tau$ (within the phase space
of the harmonic oscillator) and the 
subsequent BE the absorption images are taken.

\begin{figure}
\begin{minipage}{10cm}
\includegraphics[height=8cm,clip]{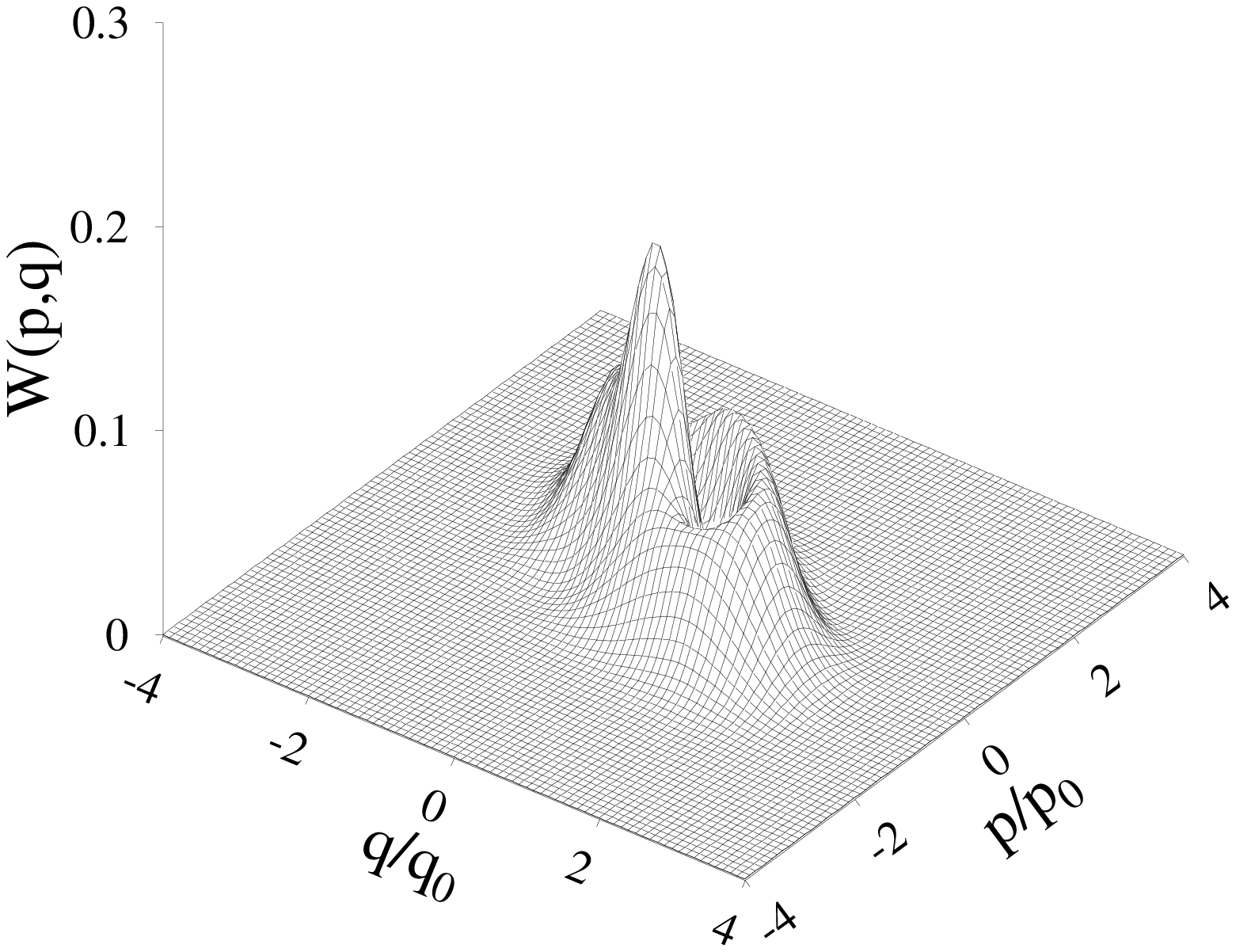}
\end{minipage}
\begin{minipage}[b]{6cm}
\includegraphics[height=5cm,clip]{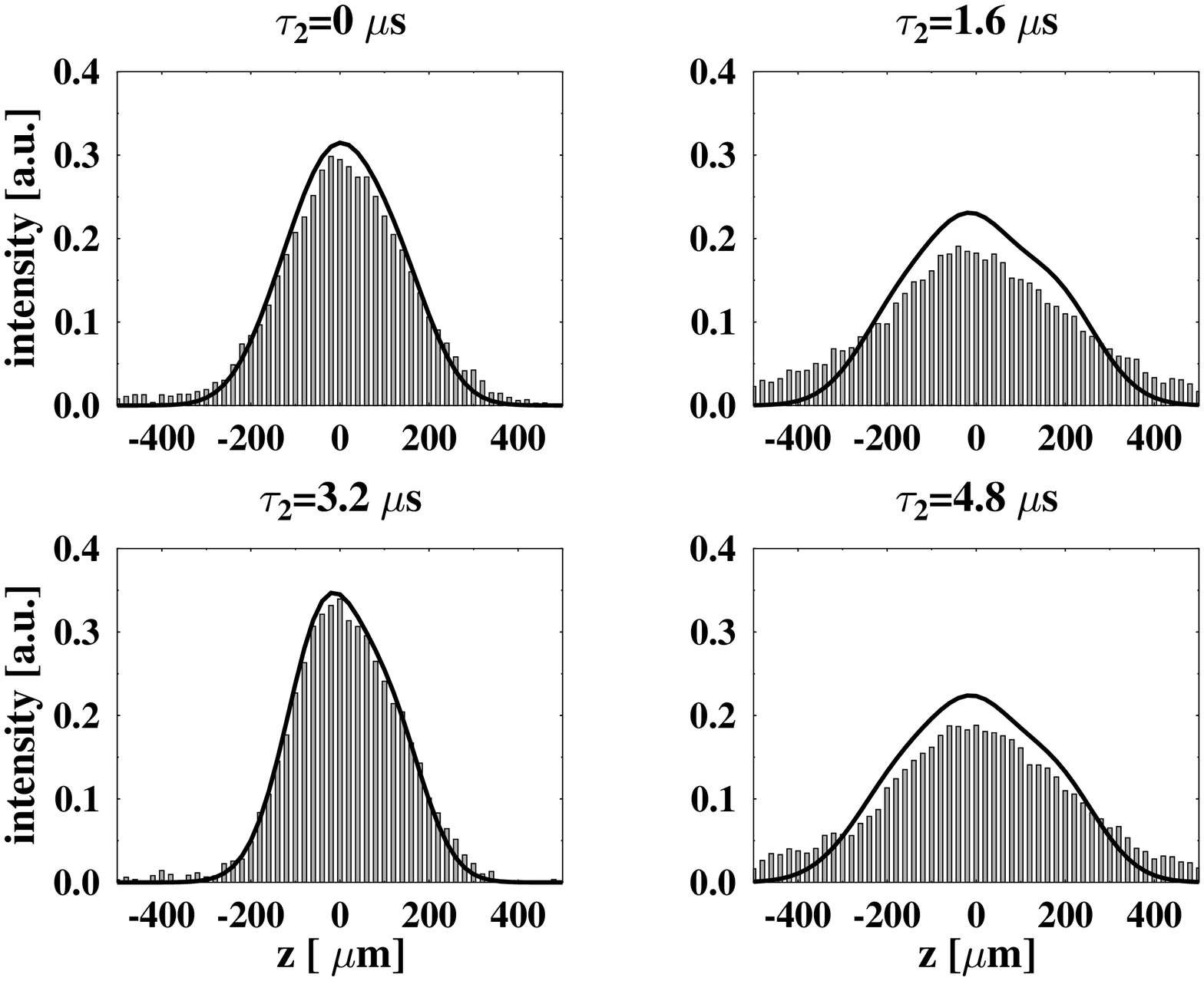}
\end{minipage}
\caption{
The  Wigner function reconstructed
from the experimental data 
obtained at the ENS, Paris.
The experimental setting is the same as for Fig.~1.
From the experimental data we have  inferred the mean number of 
motional quanta $\bar{n} \approx 1.0$, while  the reconstructed value is  
$\bar{n}^\prime \approx 1.1$. Deviation of the measured
and predicted values of observables is $\Delta F=0.09$
and entropy of the reconstructed mixture state is $S=1.0$.
}%
\end{figure}

The considered data are for the initial vacuum state which means
that under ideal conditions a squeezed state
$|\psi_1\rangle=\exp(-i t_1 \hat{p}^2/2m) |0\rangle$ should be prepared. 
 The vertical spatial distributions obtained from the measured 
2D-absorption images are discretized into pixels 
(bins) with the pixel width 5.45 $\mu$m. 
The optical density of each pixel is averaged in the horizontal direction 
in which the absorption images are divided into 50 rows, each 3.9 $\mu$m wide
(these rows cover the size of the cloud in the horizontal direction). 
For the reconstruction via the MaxEnt principle four 
 vertical spatial distributions  for ``rotation'' times 
$\tau=0, 1.6, 3.2$ and $4.8 ~\mu$s are taken. 
The selected  times
roughly correspond to rotations within the phase space by 
$\omega_z \tau=0, \pi/4, \pi/2$ and $3\pi/4$, respectively.
Unfortunately, the 
mean excitation number $\hat{n}$ for measured state $|\psi_1\rangle$ 
was not measured explicitly in the experiment, therefore we have to estimate
it as follows: 
During the free expansion 
period the rms size of the cloud increases by $\Delta x=p_0 \tau_1/m$.
The corresponding increase of the potential energy 
$\frac{1}{2} m \omega_z^2 (\Delta x)^2$ in units
$\hbar \omega_z$ gives us the increase of the number of 
excitation quanta with respect to the initial state $|\psi_0\rangle$.
For $\tau_1=4$ $\mu$ and the initial vacuum
it means $\bar{n}\approx 1$.
Experiments can be realized also for higher $\tau_1$.
For example, $\tau_1=8$ $\mu$ leads to $\bar{n}\approx 4$.
However, as mentioned above, such ``squeezed'' states 
with a significant contribution of higher phonon number states 
violate the underlying harmonic approximation for the potential.
To keep a coherent control an an-harmonic part of the potential
has to be taken into account.

The result of the  reconstruction via the MaxEnt principle 
is shown in Fig.~5.
The deviation of the fitted and measured values is
$\Delta F=0.09$ and the entropy of the reconstructed
state  $S=1.0$. It means that the reconstructed state
is a statistical 
mixture. We see a two peak structure which suggests
that there is a mixture of two squeezed states coherently displaced 
from each other. 
It is caused by the fact that the vertical center of the cloud was not 
fixed in the experiment and it has to be determined 
by our fit for each  measured BE absorption image separately.  
Assuming a priori knowledge that the Wigner function has a symmetric
shape with respect to the origin of the phase space 
(i.e. there is no coherent amplitude) a Gaussian fit can be used to 
determine the center of the cloud for each vertical  distribution.
For states with a non-zero  coherent amplitude the center of the cloud 
should be fixed already in the experiment.

It turns out that the reconstruction 
results do not describe the squeezed vacuum state as was originally
expected \cite{Salomon00}.
The main reason is that the mean phonon number was not measured 
directly in the experiment. It can be inferred only indirectly from the 
ideal case without any incoherence during preparation or measurement. 
As we discussed above, it is essential to include the information 
about the mean number of vibrational quanta
into the MaxEnt reconstruction scheme.  
In optical tomography the analogous information about mean photon number 
can be obtained from distributions of two ``orthogonal'' quadratures.
In our case it could correspond to two absorption images
such that $\omega_z (\tau_j-\tau_k)=\pi/2$. However, it would require
a precise timing of the evolution within the harmonic trapping potential.
Therefore the mean number of vibrational quanta should be determined 
in an independent measurement.

 Another problem arises from a slow convergence of anti-squeezed 
spatial distributions which are derived directly from the measured 
absorption images. 
In particular, the convergence of tails is too slow for those 
``rotations'' which corresponds to anti-squeezed phases, 
i.e. $\tau=1.6$, $4.8~\mu\mbox{s}$ (see insets of Fig.~5).
The slow convergence is reflected by the presence of non-negligible 
backgrounds for Gaussian fits to these spatial distributions. 
If we eliminate (subtract) these backgrounds from the measured 
distributions the MaxEnt reconstruction gives almost the same 
Wigner function as in Fig.~5 but with a highly reduced deviation function 
$\Delta F=0.02$ (comparing to $\Delta F=0.09$ in Fig.~5).
Such background in these absorption images can be caused by an 
incoherence associated with a violation of the 
the harmonic approximation. 
In fact, in our analysis we have neglected the change of the oscillation 
frequency along the $z$-axis. In recent experiments, the oscillation frequency 
decreases 10\% from $\omega_z$ for micro-wells 
at the edge of the initial cloud.

\section{Conclusions}
\label{sec5}

We have presented a very efficient reconstruction scheme for
the reconstruction of motional states of atoms based on 
the {\em MaxEnt} principle. The main advantage of the scheme
is that it always results in reconstructions which are physical
states (unlike in the case of the maximum likelihood estimation
which can results in nonphysical estimations).
Moreover, the scheme is very efficient in a sense that
it requires just a small number of tomographic phase-space cuts.

We have applied the scheme 
for a reconstruction of  motional quantum states
of neutral atoms. As an example have we analyzed 
the experimental data obtained by 
by the group of C. Salomon at the ENS in Paris and we reconstruct
the Wigner function of motional quantum states of Cs atoms trapped in
the optical lattice.
In our analysis we have neglected 
the change of the oscillation frequency along $z$ axis
in recent experiments.  
The dispersion of the oscillation frequency is of the order of a few percent. 
This source of errors can significantly affect the quality of 
a quantum state preparation and its reconstruction.
In addition, only up to first 10 bound states of micro-wells 
of the optical lattice can be approximated by a harmonic potential.
It implies limits on coherent manipulations of quantum states.
It means that states with a significant contribution of higher
number (Fock) 
states cannot be prepared and manipulated in a controlled way.

\acknowledgements
This work was supported
by the European Union  projects QUBITS and QUEST under the contracts
IST-1999-13021 and HPRN-CT-2000-00121, respectively.
We thank Christophe Salomon and Isabelle Bouchoule for providing
us with their experimental data and 
for helpful discussions and correspondence.



\begin{thebibliography}{xx}

\bibitem{Jaynes1963}
E.\ T. Jaynes, {\em Information theory and statistical mechanics},
in {\em 1962 Brandeis Lectures} vol. 3, ed. K. W. Ford
(Benjamin, Elmstord, New York, 1963).

\bibitem{Nielsen2000}
M.\ A.\ Nielsen and I.\ L.\ Chuang,
{\em Quantum Computation and Quantum Information}
(Cambridge University Press, Cambridge, 2000).

\bibitem{Wineland98}
For recent review articles on trapped ions see
D.J. Wineland {\sl et al.}: {\em Fortschr. Phys.} {\bf 46}, 363 (1998);
H. C. N\"agerl{\sl et al.}, {\em Fortschr. Phys.} {\bf 48}, 623 (2000);
and references therein.


\bibitem{Bouchoule99}
I. Bouchoule, H. Perrin, A. Kuhn, M. Morinaga, and C. Salomon,
  {\em Phys. Rev. A} {\bf 59}, R8 (1999).

\bibitem{Morinaga99}
M. Morinaga, I. Bouchoule, J.-C. Karam, and C. Salomon,
  {\em Phys. Rev. Lett.} {\bf 83}, 4037 (1999).


\bibitem{Jaynes1957}
E.T. Jaynes, {\em Phys. Rev.} {\bf 108}, 171 (1957);
{\em ibid.~} {\bf 108}, 620 (1957) 620;
{\em Am. J. Phys.} {\bf 31}, 66 (1963).

\bibitem{Fick1990}
 E. Fick and G. Sauermann:
{\em  The Quantum Statistics of Dynamic Processes}
(Springer Verlag, Berlin, 1990).

\bibitem{Kapur1992}
J.N.Kapur and H.K.Kesavan:
{\em Entropic Optimization Principles with Applications}
(Academic Press, New York, 1992).  

\bibitem{Newton68}
{  R.G.Newton} and {  Bing-Lin Young},
  {\em Ann. Phys.} (N.Y.) {\bf 49}, 393 (1968).

\bibitem{Katz1967}
{ A. Katz}:
  {\em Principles of Statistical  Mechanics},
  (W.H. Freeman and Company, San Francisco, 1967);
    \newline
{A. Hobson},
  {\em Concepts in Statistical  Mechanics},
  (Gordon Breach Science Publishers, New York, 1971). 

\bibitem{Buzek1996}
V. Bu\v{z}ek, G. Adam, and G. Drobn\'y, {\em Ann. Phys. (N.Y.)} {\bf 245},
37 (1996);  
 V.Bu\v{z}ek, G. Drobn\'y, R.Derka, G.Adam, and  H.Wiedeman, 
{\em Chaos, Solitons \& Fractals} {\bf 10}, 981--1074 (1999). 


\bibitem{Hradil2000}
Z. Hradil, J. Summhammer, and H. Rauch,
   {\em Phys. Lett. A} {\bf 261}, 20 (2000).

\bibitem{Vogel1989}
{  K.Vogel} and {  H.Risken},
  {\em Phys. Rev. A} {\bf 40}, 2847 (1989).

\bibitem{Leonhardt1997}
U.Leonhardt, {\em Measuring the quantum state of light}
(Cambridge University Press, Cambridge, 1997).  

\bibitem{Buzek00}
{V. Bu\v{z}ek and G. Drobn\'y}, {\em J. Mod. Optics} {\bf 47}, 2823 (2000).  


\bibitem{Smithey93}
{  D.T. Smithey, M. Beck, M.G. Raymer} and {  A. Faridani},
  {\em Phys. Rev. Lett.}{\bf 70}, 1244 (1993);
  M. Beck, D.T. Smithey, and M.G. Raymer,
  {\em Phys. Rev. A} {\bf 48}, R890 (1993);
{  M.G. Raymer, M. Beck}, and {  D.F. Mc\,Alister},
  {\em Phys. Rev. Lett.} {\bf 72}, 1137 (1994);

\bibitem{Kurtsiefer96}  S. Schiller, G. Breitenbach, S.F.
Pereira, T. M\"{u}ller, and  J. Mlynek, {\em Phys. Rev. Lett.} {\bf 77},
2933 (1996);
\newline Ch. Kurtsiefer, T. Pfau, and  J. Mlynek, {\em
Nature} {\bf 386}, 150 (1997).        

\bibitem{Wigner}
{ E.P.Wigner}, {\em Phys. Rev.} {\bf 40}, 749 (1932); see also
    \newline
{ M.Hillery, R.F.O'Connell, M.O.Scully}, and { E.P.Wigner},
  {\em Phys. Rep.} {\bf 106}, 121 (1984).   



\bibitem{Buzek1995}
V.Bu\v{z}ek and P.L.Knight: {\em
  Quantum interference, superposition states of light and non-classical
  effects},
  in {\em Progress in Optics, Vol. 34}, ed. E. Wolf (North Holland,
Amsterdam,
  1995), p.1.  

\bibitem{foot1}
We should stress that due to random fluctuations of ``observed'' mean values 
(\ref{3.17}) physically incompatible data might be obtained.  
It means that there does not
exist a physical density operator which could fit perfectly the results
of the noisy measurement (i.e. leading to $\Delta F\to 0$).
Obviously increasing the relative error of the measurement we increase 
also a number of incompatible results and the reconstruction becomes 
meaningless. 


\bibitem{Salomon00}
C. Salomon and I. Bouchoule, private communication. 

\end{thebibliography}
\end{document}